\documentclass[mathleft]{an}
\usepackage{graphicx}
\usepackage{epsfig}
\usepackage{times}
\usepackage{natbib}

\renewcommand\refname{}

\begin{document}

% The following seven commands are intended for editorial usage and should be ignored by
% the author(s).
\Pagespan{1052}{1054}% Document's page range. 
% If second parameter is left empty, the last page is computed automatically.
\Yearpublication{2011}%
\Yearsubmission{2011}%
\Month{09}%   
\Volume{332}%  
\Issue{9/10}% 
 \DOI{10.1002/asna.201111615}% 

\title{Star-planet interactions and selection effects from planet detection\\[1.5pt] methods}

\author{K. Poppenhaeger\inst{}\thanks{Corresponding author:
  {katja.poppenhaeger@hs.uni-hamburg.de}}
%Example 
%for footnote, note the usage of the \texttt{fnmsep}
%command as separator between institute number and footnote mark} 
\and  J.H.M.M. Schmitt\inst{}
}
\titlerunning{SPI selection effects}
\authorrunning{K. Poppenhaeger \& J.H.M.M. Schmitt}
\institute{
Hamburger Sternwarte, Gojenbergsweg 112, 21029 Hamburg, Germany}

\received{2011 Sep 20}
\accepted{2011 Oct 6}
\publonline{2011}


\abstract{%
Planets may have effects on their host stars by tidal or magnetic interaction. Such star-planet interactions are thought to enhance the activity level of the host star. However, stellar activity also affects the sensitivity of planet detection methods. Samples of planet-hosting stars which are investigated for such star-planet interactions are therefore subject to strong selection effects which need to be taken into account.}

\maketitle

\section{Introduction}

Magnetic activity is a well-known phenomenon in the Sun, and it is ubiquitous among late-type stars. The magnetic field of cool stars is highly structured at a local scale and is mainly responsible for the stellar chromospheric and coronal emission. Close binary stars display a much higher activity level than single stars of a comparable age and spectral type. This is mainly due to the fast stellar rotation in binaries which is preserved by tidal locking, but there is also evidence for direct magnetic interaction and coronal emission between the two components of close binaries \citep{Siarkowski1996}. Regarding a star and its giant close-in planet as a binary system with a very small mass ratio, such interactions might also be expected in star-planet systems. 

The theoretical framework of star-planet interactions  (SPI) discerns between
tidal and magnetic interaction,  which are both thought to be able to
increase the stellar activity level \citep{Cuntz2000}. Tidal bulges caused by
the planet in the stellar atmosphere rise and subside with respect to the
stellar rotational frame if the planetary orbit and stellar rotation are not yet
synchronized. This may increase the turbulence in the stellar photosphere and
lead to faster entanglement of footpoints of coronal loops and therefore more
frequent flaring. Magnetic interactions of the planet on the star may arise from
reconnection events between stellar and planetary magnetic field lines \citep{Ip2004}, the propagation of Alfv\'{e}n waves in the stellar wind \citep{Preusse2006}, or from disturbing effects that the planetary magnetosphere may have on stellar coronal loops such as flare triggering \citep{Pillitteri2010}.

\section{SPI signatures in stellar coronae and chromospheres}

Observing activity signatures caused by SPI is a difficult task since all cool
stars display some level of intrinsic variability, causing both short-term and
long-term activity  changes. However, in some individual star-planet systems,
indications for such signatures have been found. Out of ten program stars with
Hot Jupiters \citep{Shkolnik2005}, the two stars HD\,179949 and $\upsilon$\,And
showed variations in the activity-influenced chromospheric Ca\,{\sc ii}~K line emission
 which were in phase with the planetary orbit, not the stellar rotation period.
 However, subsequent observations showed that the chromospheric variability had
 switched back to the stellar rotation period again \citep{Shkolnik2008, PoppenhaegerLenz2011}. Recently, a possible influence of a Hot Jupiter on the photospheric spot distribution of its host star, CoRoT-6, was reported \citep{Lanza2011}. For the star HD~189733, which hosts a transiting Hot Jupiter, an X-ray flare has been detected which occurred directly after the eclipse (not the transit) of the planet; the authors interpret this as active region flares triggered by the planet \citep{Pillitteri2010}.

Several analyses of samples of planet-hosting stars have been conducted as well.
In an initial statistical study which tested for trends of the stellar X-ray
luminosity with planetary semimajor axis, \cite{Kashyap2008} reported elevated
X-ray activity in stars with close-in planets. However, this trend could not be
recovered in a later study using a complete sample of planet-hosting stars in
the solar neighborhood \citep{Poppenhaeger2010}. A study of the chromospheric activity of planet-hosting stars did not detect a dependence of stellar activity on semimajor axis or planetary mass \citep{CantoMartins2011}.

\section{The influence of selection effects from planet-detection methods}

Samples of planet-hosting stars can be a powerful tool to identify SPI-related
trends in coronal activity. However,  strong biases can arise from
planet-detection methods. It is well known that stellar activity can mask radial
velocity (RV) signals induced by planets, so that detecting small planets around
active stars is notoriously difficult. In the light of observational programs
which search for rocky  planets in habitable zones, M~dwarfs as host stars have 
gained lots of attention recently. Detailed studies of spot coverages and their
influence on the RV detectability of  rocky planets have been conducted
\citep{Dumusque2011, BarnesJeffers2011}. In transit surveys, stellar activity has an influence on the derived planetary parameters, as star spots which are covered by the planet during the transit distort the transit profile and can lead to smaller estimates for the planetary radius if not accounted for \citep{Czesla2009}.

\begin{figure}%[ht!]
\begin{center}
\epsfig{width=0.485\textwidth,file=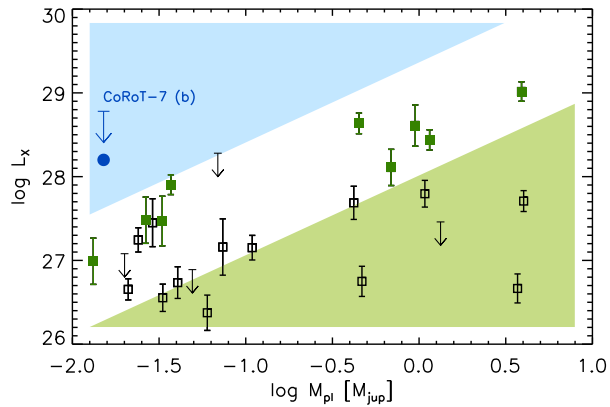}
\caption{(online colour at: www.an-journal.org) X-ray luminosities of stars with close-in planets (${a_{\rm pl}<0.15}$~AU); flux-limited subsample as green filled squares, volume-limited complete sample in green and black, estimated $L_{\rm X}$ of CoRoT-7 (which hosts a transit-detected super-earth) as blue filled circle.}
\label{fig}
\end{center}
\end{figure}

Other biases concerning the stellar activity level can  arise from flux-limited
surveys in X-rays. For example the X-ray detections of planet-hosting stars from
the ROSAT All-Sky Survey (RASS), which is strongly flux limited,  show a very
prominent trend in stars with close-in  ($<$0.15~AU) planets of stellar X-ray
luminosity with the planetary mass \citep{Scharf2010}, depicted by green data points
in Fig.~\ref{fig}. However, the RASS is only complete with regard to stellar
X-ray detections out to 5--10 pc, depending on the spectral type, so there is a distance-related selection effect in this set of data.

Therefore, we composed a sample of all known planet-hosting stars within a
distance of 30 pc from the Sun \citep{Poppenhaeger2011}. Combining new and archival X-ray observations conducted with XMM-Newton and ROSAT  for this, our sample is practically complete with regard to X-ray detections (52 of 72 stars detected). The main difference to the RASS sample is that there are many stars with low $L_{\rm X}$ which harbour massive planets, filling up the lower right corner of the diagram in Fig.~\ref{fig} (shaded in green). The previous lack in such systems obviously stems from the X-ray flux limit of the RASS data. This means that there is no dependence of some minimal $L_{\rm X}$ on the planetary mass. 

However, there are no stars with high $L_{\rm X}$ and small planets in the sample (upper left corner, shaded in blue).  Again, this is not an SPI signature, but an effect of the non-completeness with regard to planet detections: In the solar neighborhood, planets are mostly detected by radial velocity (RV) shifts. Stellar activity makes the detection of RV shifts more difficult, so that for active stars only strong RV signals can be detected, requiring a heavy planet, low-mass host star, or both. This means that small planets are only detected around low-activity stars by the RV method.

Transit detections are less influenced by stellar activity, and several small transiting planets have been discovered orbiting quite active stars. As a prominent example we discuss here the case of CoRoT-7b. CoRoT-7b was first detected as a companion causing a 0.034\,\% flux reduction of its host star during transit \citep{Leger2009}. The exact mass determination with the radial velocity method proved to be difficult, as the host star displays considerable magnetic activity, and the expected RV signal was of the order of only a few m/s. Detailed analyses have pinned down the planetary mass now to 6.9\,M$_{\oplus}$, classifying CoRoT-7b as a super-earth \citep{Hatzes2010}. 

The host star CoRoT-7 is a main-sequence star of spectral type K0. Its X-ray
flux has not been detected in the ROSAT All-Sky Survey (RASS); this
non-detection yields on upper limit to CoRoT-7's X-ray luminosity of $\log L_{\rm X} <
28.78$, assuming a distance of $\approx$150~pc \citep{Leger2009}. The
chromospheric activity level inferred from the chromospheric Ca\,{\sc ii} H and K lines
is given by \cite{Queloz2009} to be $\log R_{HK}' = -4.61$. Using the relation
of chromospheric activity $\log R_{HK}'$ and coronal activity indicator $\log
L_{\rm X}/L_{\rm bol}$ \citep{Mamajek2008}, we estimate \hbox{CoRoT-7's}  coronal activity indicator to be ${\approx\! -5.1}$. With bolometric corrections from Flower (1996), we derive a bolometric luminosity of $\log L_{\rm bol}=33.3$ and an estimated X-ray luminosity of $\log L_{\rm X}=28.2$. This is compatible with the upper limit derived from the RASS non-detection. The expected X-ray luminosity for CoRoT-7 is about one order of magnitude higher than for other (RV-detected) planet host stars with very small planets, as is evident from Fig.~\ref{fig}.

This prominent example shows that the absence of detected low-mass planets around active stars is a selection effect. This effect is very pronounced in near-by planet-hosting stars, because the radial velocity technique is the main planet detection method in the solar neighborhood. Transit-search missions such as Kepler and CoRoT are able to detect small planets around more active stars, but they mostly detect planets around distant (${d>100}$~pc) stars. In principle, activity studies of stars with transit-detected planets can provide a new angle to the question of planet-induced activity features. However, the large distance to the host stars makes the characterization of their activity level via optical spectroscopy or X-ray flux detections  time-expensive. A tho\-rough treatment of selection effects in RV-detected samples is therefore inevitable.

\section{Conclusion}

We have investigated trends in the X-ray emission of planet-hosting stars within a distance of 30~pc from the Sun. We found that the apparent trend of X-ray luminosity with planetary mass can be explained by observational biases. Our analysis shows that selection effects which are introduced by the planet detection methods are crucial in the analysis of SPI signatures as they can produce spurious trends. In comparison, samples with transit-detected planets may yield more insight, since the transit method is better suited to find small planets around active stars than the radial velocity method.

\bibliographystyle{aa}
\bibliography{../../Docs/katjasbib.bib}

\begin{thebibliography}{21}
\expandafter\ifx\csname natexlab\endcsname\relax\def\natexlab#1{#1}\fi

\bibitem[{{Barnes} {et~al.}(2011){Barnes}, {Jeffers}, \&
  {Jones}}]{BarnesJeffers2011}
{Barnes}, J.~R., {Jeffers}, S.~V., \& {Jones}, H.~R.~A. 2011, \mnras, 412, 1599

\bibitem[{{Canto Martins} {et~al.}(2011){Canto Martins}, {Das Chagas}, {Alves},
  {Le{\~a}o}, {de Souza Neto}, \& {de Medeiros}}]{CantoMartins2011}
{Canto Martins}, B.~L., {Das Chagas}, M.~L., {Alves}, S., {et~al.} 2011, \aap,
  530, A73

\bibitem[{{Cuntz} {et~al.}(2000){Cuntz}, {Saar}, \& {Musielak}}]{Cuntz2000}
{Cuntz}, M., {Saar}, S.~H., \& {Musielak}, Z.~E. 2000, \apjl, 533, L151

\bibitem[{{Czesla} {et~al.}(2009){Czesla}, {Huber}, {Wolter}, {Schr{\"o}ter},
  \& {Schmitt}}]{Czesla2009}
{Czesla}, S., {Huber}, K.~F., {Wolter}, U., {Schr{\"o}ter}, S., \& {Schmitt},
  J.~H.~M.~M. 2009, \aap, 505, 1277

\bibitem[{{Dumusque} {et~al.}(2011){Dumusque}, {Santos}, {Udry}, {Lovis}, \&
  {Bonfils}}]{Dumusque2011}
{Dumusque}, X., {Santos}, N.~C., {Udry}, S., {Lovis}, C., \& {Bonfils}, X.
  2011, \aap, 527, A82

\bibitem[{{Hatzes} {et~al.}(2010){Hatzes}, {Dvorak}, {Wuchterl}, {Guterman},
  {Hartmann}, {Fridlund}, {Gandolfi}, {Guenther}, \&
  {P{\"a}tzold}}]{Hatzes2010}
{Hatzes}, A.~P., {Dvorak}, R., {Wuchterl}, G., {et~al.} 2010, \aap, 520, A93

\bibitem[{{Ip} {et~al.}(2004){Ip}, {Kopp}, \& {Hu}}]{Ip2004}
{Ip}, W.-H., {Kopp}, A., \& {Hu}, J.-H. 2004, \apjl, 602, L53

\bibitem[{{Kashyap} {et~al.}(2008){Kashyap}, {Drake}, \& {Saar}}]{Kashyap2008}
{Kashyap}, V.~L., {Drake}, J.~J., \& {Saar}, S.~H. 2008, \apj, 687, 1339

\bibitem[{{Lanza} {et~al.}(2011){Lanza}, {Bonomo}, {Pagano}, {Leto}, {Messina},
  {Cutispoto}, {Moutou}, {Aigrain}, {Alonso}, {Barge}, {Deleuil}, {Fridlund},
  {Silva-Valio}, {Auvergne}, {Baglin}, \& {Collier Cameron}}]{Lanza2011}
{Lanza}, A.~F., {Bonomo}, A.~S., {Pagano}, I., {et~al.} 2011, \aap, 525, A14

\bibitem[{{L{\'e}ger} {et~al.}(2009){L{\'e}ger}, {Rouan}, {Schneider}, {Barge},
  {Fridlund}, {Samuel}, {Ollivier}, {Guenther}, {Deleuil}, {Deeg}, {Auvergne},
  {Alonso}, {Aigrain}, {Alapini}, {Almenara}, {Baglin}, {Barbieri}, {Bruntt},
  {Bord{\'e}}, {Bouchy}, {Cabrera}, {Catala}, {Carone}, {Carpano}, {Csizmadia},
  {Dvorak}, {Erikson}, {Ferraz-Mello}, {Foing}, {Fressin}, {Gandolfi},
  {Gillon}, {Gondoin}, {Grasset}, {Guillot}, {Hatzes}, {H{\'e}brard}, {Jorda},
  {Lammer}, {Llebaria}, {Loeillet}, {Mayor}, {Mazeh}, {Moutou}, {P{\"a}tzold},
  {Pont}, {Queloz}, {Rauer}, {Renner}, {Samadi}, {Shporer}, {Sotin}, {Tingley},
  {Wuchterl}, {Adda}, {Agogu}, {Appourchaux}, {Ballans}, {Baron}, {Beaufort},
  {Bellenger}, {Berlin}, {Bernardi}, {Blouin}, {Baudin}, {Bodin}, {Boisnard},
  {Boit}, {Bonneau}, {Borzeix}, {Briet}, {Buey}, {Butler}, {Cailleau},
  {Cautain}, {Chabaud}, {Chaintreuil}, {Chiavassa}, {Costes}, {Cuna Parrho},
  {de Oliveira Fialho}, {Decaudin}, {Defise}, {Djalal}, {Epstein}, {Exil},
  {Faur{\'e}}, {Fenouillet}, {Gaboriaud}, {Gallic}, {Gamet}, {Gavalda},
  {Grolleau}, {Gruneisen}, {Gueguen}, {Guis}, {Guivarc'h}, {Guterman},
  {Hallouard}, {Hasiba}, {Heuripeau}, {Huntzinger}, {Hustaix}, {Imad},
  {Imbert}, {Johlander}, {Jouret}, {Journoud}, {Karioty}, {Kerjean},
  {Lafaille}, {Lafond}, {Lam-Trong}, {Landiech}, {Lapeyrere}, {Larqu{\'e}},
  {Laudet}, {Lautier}, {Lecann}, {Lefevre}, {Leruyet}, {Levacher}, {Magnan},
  {Mazy}, {Mertens}, {Mesnager}, {Meunier}, {Michel}, {Monjoin}, {Naudet},
  {Nguyen-Kim}, {Orcesi}, {Ottacher}, {Perez}, {Peter}, {Plasson}, {Plesseria},
  {Pontet}, {Pradines}, {Quentin}, {Reynaud}, {Rolland}, {Rollenhagen},
  {Romagnan}, {Russ}, {Schmidt}, {Schwartz}, {Sebbag}, {Sedes}, {Smit},
  {Steller}, {Sunter}, {Surace}, {Tello}, {Tiph{\`e}ne}, {Toulouse}, {Ulmer},
  {Vandermarcq}, {Vergnault}, {Vuillemin}, \& {Zanatta}}]{Leger2009}
{L{\'e}ger}, A., {Rouan}, D., {Schneider}, J., {et~al.} 2009, \aap, 506, 287

\bibitem[{{Mamajek} \& {Hillenbrand}(2008)}]{Mamajek2008}
{Mamajek}, E.~E. \& {Hillenbrand}, L.~A. 2008, \apj, 687, 1264

\bibitem[{{Pillitteri} {et~al.}(2010){Pillitteri}, {Wolk}, {Cohen}, {Kashyap},
  {Knutson}, {Lisse}, \& {Henry}}]{Pillitteri2010}
{Pillitteri}, I., {Wolk}, S.~J., {Cohen}, O., {et~al.} 2010, \apj, 722, 1216

\bibitem[{{Poppenhaeger} {et~al.}(2011){Poppenhaeger}, {Lenz}, {Reiners},
  {Schmitt}, \& {Shkolnik}}]{PoppenhaegerLenz2011}
{Poppenhaeger}, K., {Lenz}, L.~F., {Reiners}, A., {Schmitt}, J.~H.~M.~M., \&
  {Shkolnik}, E. 2011, \aap, 528, A58+

\bibitem[{{Poppenhaeger} {et~al.}(2010){Poppenhaeger}, {Robrade}, \&
  {Schmitt}}]{Poppenhaeger2010}
{Poppenhaeger}, K., {Robrade}, J., \& {Schmitt}, J.~H.~M.~M. 2010, \aap, 515,
  A98+

\bibitem[{{Poppenhaeger} \& {Schmitt}(2011)}]{Poppenhaeger2011}
{Poppenhaeger}, K. \& {Schmitt}, J.~H.~M.~M. 2011, \apj, 735, 59

\bibitem[{{Preusse} {et~al.}(2006){Preusse}, {Kopp}, {B{\"u}chner}, \&
  {Motschmann}}]{Preusse2006}
{Preusse}, S., {Kopp}, A., {B{\"u}chner}, J., \& {Motschmann}, U. 2006, \aap,
  460, 317

\bibitem[{{Queloz} {et~al.}(2009){Queloz}, {Bouchy}, {Moutou}, {Hatzes},
  {H{\'e}brard}, {Alonso}, {Auvergne}, {Baglin}, {Barbieri}, {Barge}, {Benz},
  {Bord{\'e}}, {Deeg}, {Deleuil}, {Dvorak}, {Erikson}, {Ferraz Mello},
  {Fridlund}, {Gandolfi}, {Gillon}, {Guenther}, {Guillot}, {Jorda}, {Hartmann},
  {Lammer}, {L{\'e}ger}, {Llebaria}, {Lovis}, {Magain}, {Mayor}, {Mazeh},
  {Ollivier}, {P{\"a}tzold}, {Pepe}, {Rauer}, {Rouan}, {Schneider},
  {Segransan}, {Udry}, \& {Wuchterl}}]{Queloz2009}
{Queloz}, D., {Bouchy}, F., {Moutou}, C., {et~al.} 2009, \aap, 506, 303

\bibitem[{{Scharf}(2010)}]{Scharf2010}
{Scharf}, C.~A. 2010, \apj, 722, 1547

\bibitem[{{Shkolnik} {et~al.}(2008){Shkolnik}, {Bohlender}, {Walker}, \&
  {Collier Cameron}}]{Shkolnik2008}
{Shkolnik}, E., {Bohlender}, D.~A., {Walker}, G.~A.~H., \& {Collier Cameron},
  A. 2008, \apj, 676, 628

\bibitem[{{Shkolnik} {et~al.}(2005){Shkolnik}, {Walker}, {Bohlender}, {Gu}, \&
  {K{\"u}rster}}]{Shkolnik2005}
{Shkolnik}, E., {Walker}, G.~A.~H., {Bohlender}, D.~A., {Gu}, P., \&
  {K{\"u}rster}, M. 2005, \apj, 622, 1075

\bibitem[{{Siarkowski} {et~al.}(1996){Siarkowski}, {Pres}, {Drake}, {White}, \&
  {Singh}}]{Siarkowski1996}
{Siarkowski}, M., {Pres}, P., {Drake}, S.~A., {White}, N.~E., \& {Singh}, K.~P.
  1996, \apj, 473, 470

\end{thebibliography}

\end{document}